\documentclass[12pt]{article}
\usepackage[left=0.7in,right=0.7in,top=1in,bottom=1in,a4paper]{geometry}

\usepackage{setspace,amsmath,latexsym,cite,amssymb,epsfig,amsfonts,subfig}
\usepackage{url}
\usepackage{graphicx}
\usepackage{psfrag}
\usepackage{footmisc}
\usepackage{multirow}
\usepackage{color}
\usepackage{soul}
\usepackage{multicol}
\usepackage{mathtools}
\usepackage{algorithm}
\usepackage[noend]{algpseudocode}
\definecolor{asparagus}{rgb}{0, 0.5, 0.0}
\DeclareMathAlphabet{\mymathbb}{U}{BOONDOX-ds}{m}{n}
\usepackage{kotex}
\usepackage{comment}
\usepackage{authblk}
\usepackage{etoolbox}
\usepackage{nth}

\graphicspath{{.}{../eps/}{./images/}{./images/result/}}
\newcommand*\diff{\mathop{}\!\mathrm{d}}

\usepackage{amsthm}

\usepackage[cmintegrals]{newtxmath}
\usepackage{siunitx}

\usepackage{cuted}
\usepackage[font=small, labelsep=period]{caption}
\usepackage{makecell}
\usepackage{tabu}
\usepackage{bbm}

\makeatletter
\patchcmd{\@maketitle}{\raggedright}{\centering}{}{}
\patchcmd{\@maketitle}{\raggedright}{\centering}{}{}
\makeatother

\title{Electrophoretic Molecular Communication with Time-Varying Electric Fields}
\author[1]{Sunghwan~Cho\thanks{sunghwan.cho@eng.ox.ac.uk}}
\author[1]{Thomas~C.~Sykes\thanks{thomas.sykes@eng.ox.ac.uk/ t.c.sykes@outlook.com}}
\author[1]{Justin~P.~Coon\thanks{justin.coon@eng.ox.ac.uk}}
\author[1]{Alfonso~A.~Castrej\'{o}n-Pita\thanks{alfonso.castrejon-pita@wadham.ox.ac.uk}}

\affil[1]{Department of Engineering Science, University of Oxford,\protect\\ Oxford, OX1 3PJ, UK}

\begin{document}

\maketitle

\abstract
This article investigates a novel electrophoretic molecular communication mechanism that utilizes a time-varying electric field, which induces time-varying molecule velocities and in turn improves communication performance. For a sinusoidal field, we specify favorable signal parameters (e.g., phase and frequency) that yield excellent communication link performance.  We also analytically derive an optimized field function by formulating an appropriate cost function and solving the Euler-Lagrange equation.  In our setup, the field strength is proportional to the molecular velocity; we verify this assumption by solving the Basset-Boussinesq-Oseen equation for a given time-varying electric field (forcing function) and examining its implications for practical physical parameterizations of the system. Our analysis and Monte-Carlo simulation results demonstrate that the proposed time-varying approach can significantly increase the number of information-carrying molecules expected to be observed at the receiver and reduce the bit-error probability compared to the constant field benchmark.\\


\newpage

\section{Introduction}

{In the last decade, there have been considerable advancements in the fields of nanonetworks, consisting of nano-scale functional components that can perform very simple and specific tasks such as sensing, actuation, computing, and data storing~\cite{16N.F}. The interconnection of these \emph{nanomachines} allows individual components' limitations to be overcome in nanonetworks, thus expanding nanomachines' capabilities by providing them with a way to cooperate and share information. The resulting nanonetworks' potential applications are vast and varied, including industrial and consumer goods, environmental applications, biomedical sciences, and defence~\cite{08I.A}. The key challenge in exploiting nanonetworks in such applications is introducing an effective communication mechanism between the constituent nanomachines. However, there are emerging applications where conventional radio frequency communication technologies are unsafe or impractical, so there is a requirement to explore alternative communication mediums (including optical, acoustic and mechanical) or to define entirely new paradigms, such as those inspired by biology~\cite{11A.K}.

In nature, various cells and living organisms exchange information employing molecular communication (MC); that is, they use molecules as biochemical signals to encode, transmit and receive information \cite{90D.R,Leef2013}. For example, hormones can transmit signals within multicellular organisms, whereas pheromones can be secreted to communicate with members of the same species \cite{Hiyama2010}. Motivated by natural MC systems refined by evolution over countless millennia, biomimetic engineered applications exploiting MC have gained increasing attention within the communications research community as a potential communication solution for nanonetworks, kick-starting highly-interdisciplinary research in this area~\cite{16N.F}. Additional advantages of MC is that it is naturally biocompatible and consumes very little energy, both features that are expected to allow MC to be utilized in medical technologies \cite{Nakano2011,Farsad2013}.

At present, technical harnessing of molecular signaling in a fluid medium can be achieved in an engineered manner by exploiting the advection and diffusion of information-carrying molecules~\cite{12S.K}. Diffusion is the random migration by the thermal energy of molecules suspended in a fluid medium due to their collisions with other molecules in the fluid, whereas advection is a fundamental mechanism for solute particle transport in a fluid environment. Transport by advection can be categorized as force-induced drift, where advection can be induced by external forces acting on the information-carrying molecules (not on the fluid molecules), and bulk flow, where the movement of the fluid induces molecule movement. Without advection (i.e., in the purely diffusive environment), a signal distortion in which one symbol interferes with subsequent symbols, called intersymbol interference (ISI), is a significant problem in MC, especially when the distance that molecules must travel is large. Hence, with only diffusion, only low rates of transmission are generally achievable \cite{Hiyama2010}. Therefore, a means of molecule advection is often introduced too, which aims to mitigate ISI in a communication sense by migrating residual molecules away from the receiver. Consequently, exploiting both advection and diffusion can improve the potential rate of information flow in MC systems.

On the one hand, existing literature concerning MC has generally assumed the simplest case of steady and uniform advection, where `steady' implies that the flow (i.e. velocity components) is not a function of time. In contrast, `uniform' means that the flow velocity is identical throughout the environment of interest. Under the steady and uniform advection assumption, various components of MC systems have been rigorously investigated~\cite{11D.M, 12K.S, 14A.N3, 14N.K, 18V.J, 19I.A}.

On the other hand, utilizing bulk flow to propagate information-carrying molecules through a fluid medium to an intended receiver may not be suitable for applications where inducing a flow of the medium itself is problematic, unwanted, or even impossible (e.g., in lab-on-a-chip applications employing microfluidics with highly parallel arrays of reactors in which inlet and outlet ports must be shared among many chip components~\cite{14T.V}). Utilizing electric fields to controllably propagate information-carrying molecules is a potential method to resolve this limitation, rather than relying on fluid movement. The motion of dispersed particles relative to a fluid under the influence of an electric field is known as electrophoresis \cite{18H.B}, which can be readily achieved in lab-on-a-chip devices \cite{Sedgwick2008}.

Electric fields offer a degree of freedom that is inherently separate from molecular systems. High field strengths can be used to propagate information very quickly (low delay) and vice versa. This advantage leads us to contemplate that a time-varying electric field, which induces a controllable time-varying flow of information-carrying molecules, could provide a desirable improvement in communication performance. In this work, we propose a framework for utilizing such time-varying flows to achieve enhanced communication in engineered MC systems, assessing both its feasibility and communication performance. To the best of our knowledge, this work is the first to consider time-varying molecule flows induced by an electrophoresis for MC systems. The contribution of this paper can be summarized comprehensively as follows:

\begin{itemize}
    \item alongside examining the physical mechanism of molecules' motion subject to time-varying electric fields in general, we investigate a specific sinusoidal field and propose a method to choose its signal parameters in order to increase the expected number of information-carrying molecules within the receiver sphere when sampling occurs for a given bit interval;
    \item we analytically derive an optimized field (with exponential functional form) that minimises the root-mean-square error between the receiver location and the center of the transmitted molecule group, subject to a constraint on the average power of the electrical field;
    \item we show that the proposed electrophoresis framework is feasible from a fluid dynamics perspective by analyzing the effects of viscosity and mass on molecular motion in response to a time-varying electric field.
\end{itemize}

The rest of this paper is organized as follows. Section~\ref{sec:2} begins with the system model describing the transmitter and receiver, the receiver signal model, and the detection scheme in MC systems. In Section~\ref{sec:3}, the sinusoidal and optimized fields are investigated in order to increase the expected number of molecules observed by the receiver. Section~\ref{sec:4} explores the fluid dynamics underpinning electrophoretic molecular communication. Section~\ref{sec:5} gives numerical and simulation results that support our analysis. Section~\ref{sec:6} offers thoughts on future directions. Section~\ref{sec:7} concludes the paper.
}

\section{System Model}
\label{sec:2}

In this section, we introduce the system model that is subsequently used to assess the efficacy of electrophoretic MC.

\begin{figure} [t]
  \centering
  \centerline{\includegraphics[scale=0.7]{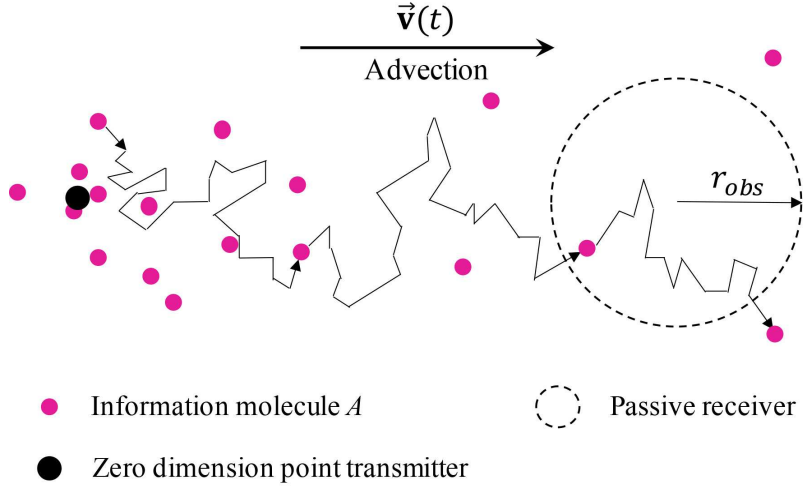}}
    \caption{Schematic description of the system model, including a point transmitter, an infinite, three-dimensional channel, and a passive receiver sphere.} 
      \label{fig:1}
\end{figure}

\subsection{Transmitter and Receiver}

The system model considered in this work is described in Fig.~\ref{fig:1}. The receiver is a sphere with radius $r_{obs}$ and volume $V_{obs}$ that is fixed and centered at the origin (i.e. $\{0, 0, 0\}$) of {an infinite, three-dimensional fluid environment of constant uniform temperature and viscosity}. The receiver is a passive observer that does not impede the migration of molecules or initiate chemical reactions. The transmitter is a point source of information molecules (called $A$ molecules) and fixed at $\{-x_0, 0, 0\}$. In addition to the transmitter, we assume that the environment has other sources of $A$ molecules, either via interference from other communication links or via some other chemical process that generates $A$ molecules. We assume that these unintended noise and interference (in a communication sense) can be characterized as a Poisson random variable with a time-varying mean.

The transmitter has a $B$-bit binary sequence $\mathbf{W} = \{ W[1], W[2], \ldots, W[B]\}$ to send to the receiver, where $W[j]$ is the $j$th information bit and $\text{Pr}(W[j]=1)=P_1$. The transmitter uses binary modulation (i.e., only two symbols $0$ and $1$ are utilized) and transmission intervals of duration $T_{\text{int}}$ seconds. To send a binary 1, $N_{EM}$ molecules are released in an impulsive manner at the start of the bit interval to mitigate ISI. To send a binary 0, no molecules are released.

Moreover, we assume that a time-varying electric field $\vec{E}(t)$ is applied uniformly over the entire environment. This field induces an electrophoretic force $\vec{F}_E(t)=q_A \vec{E}(t)$, which produces a flow of $A$ molecules with time-varying velocity $\vec{v}(t)$, where $q_A$ denotes the electric charge on a single $A$ molecule. We assume that the molecule velocity is linearly related to the electrophoretic force, i.e., $\vec{v}(t) \propto \vec{E}(t)$. The feasibility of this assumption will be validated in Section~\ref{sec:4}. Also, we define $\vec{v}(t)$ by its velocity component along each dimension, i.e. $\vec{v}(t)=\{v_x (t), v_y (t), v_z (t)\}$. The placement of the transmitter is such that $v_x (t)$ is positive in the direction of the receiver from the transmitter.In addition, electrostatic repulsion between the like-charged molecules is not taken into account in this initial study.

\subsection{Receiver Signal}
\label{sec:2.B}

The concentration of $A$ molecules (transmitted at time $t_0$) at the point defined by vector $\vec{r}$ and at time $t$ in molecule$\cdot \text{m}^{-3}$ is denoted by $C_A(\vec{r},t; t_0)$ (or written as $C_A$ for compactness). We assume that either these molecules travel independently once they are released either by the transmitter or sources of noise. In addition, due to the constant uniform temperature and viscosity of the environment, the $A$ molecules diffuse with constant diffusion coefficient $D_A$ (m$^2$/s). The differential equation describing the motion of $A$ molecules due to both advection and diffusion (via Fick's second law~\cite{18H.B}) is
\begin{equation} \label{eq:1}
\frac{\partial C_A}{\partial t} + \vec{v}(t)\cdot \nabla C_A = D_A \nabla^2 C_A.
\end{equation}
$C_A$ can be interpreted as the expected point concentration due to an emission of $N_{EM}$ molecules. By utilizing a moving reference frame {with the initial condition (IC) $C(\vec{r},t_0;t_0)=N_{EM}\delta(\vec{r}-\vec{r}_{TX})$ and boundary condition (BC) $C(\vec{r}\rightarrow \infty,t;t_0) =0$,} the expected concentration at point $\{x, y, z\}$ $\text{for} ~ t \geq t_0$ is 
\begin{equation}  \label{eq:2}
C_A(\vec{r},t; t_0)= \frac{N_{EM}}{(4 \pi D_A (t-t_0))^{3/2}} \exp\left( -\frac{\left|\vec{r}\right|^2}{4D_A (t-t_0)}\right) 
\end{equation}
where
\begin{equation} \label{eq:3}
\left|\vec{r}\right|^2=\left(x+x_0-\int_{t_0}^t v_x(t) dt \right)^2 + \left(y-\int_{t_0}^t v_y(t) dt \right)^2+ \left(z-\int_{t_0}^t v_z(t) dt \right)^2
\end{equation}
is the square of the \emph{effective} distance from the transmitter at $\{-x_0, 0, 0\}$ to $\{x, y, z\}$.

The receiver is a passive observer, so the expected number of $A$ molecules within the receiver volume {(due to a single emission of molecules)} is found by integrating \eqref{eq:2} over $V_{obs}$, which in spherical coordinates means
\begin{equation} \label{eq:4}
\overline{N_{A_0}}(t; t_0)=\int_0^{r_{obs}} \int_0^{2\pi} \int_0^\pi  C_A (\vec{r},t; t_0) r^2 \sin \theta \diff \theta \diff \phi \diff r.
\end{equation}
This integral can be simplified by utilizing the uniform concentration assumption, which assumes that the expected concentration throughout the receiver is equal to that expected at the center of the receiver, to 
\begin{equation} \label{eq:5}
\overline{N_{A_0}}(t;t_0)= V_{obs} C_A (\vec{r}_{\text{eff}},t; t_0),
\end{equation}
where {$V_{obs}=4\pi r_{obs}^3 /3$} and 
\begin{equation}\label{eq:6}
\left|\vec{r}_{\text{eff}}\right|= \sqrt{\left(x_0-\int_{t_0}^t v_x(t) dt \right)^2 + \left(\int_{t_0}^t v_y(t) dt \right)^2 + \left(\int_{t_0}^t v_z(t) dt \right)^2} 
\end{equation}
is the effective distance between the transmitter and the center of the receiver.

The statistics of the general receiver signal $N_{A_{obs}}(t)$ can be derived based on $\overline{N_{A_0}}(t; t_0)$ and the transmitted binary sequence $\mathbf{W}$, yielding the number of observed molecules due to sequential transmissions from the transmitter and noise. Assuming that $A$ molecules diffuse entirely randomly and independently, $N_{A_{obs}}(t)$ is a sum of time-varying Poisson random variables as shown in~\cite{14A.N2} with time-varying mean
\begin{equation} \label{eq:7}
\overline{N_{A_{obs}}}(t)= \overline{N_{A_{TX}}}(t)+\overline{N_{A_{n}}}(t).
\end{equation}
Here, $\overline{N_{A_{n}}}(t)$ is the mean number of molecules from the noise sources, and $\overline{N_{A_{TX}}}(t)$ is the mean number of observed molecules due to {sequential} emissions by the transmitter, i.e.,
\begin{equation} \label{eq:8}
\overline{N_{A_{TX}}}(t) = \sum_{j=1}^{\lfloor \frac{t}{T_{\text{int}}}+1 \rfloor} W[j] \overline{N_{A_0}}\left(t; (j-1)T_{\text{int}}\right).
\end{equation}

\subsection{Weighted Sum Detectors}
\label{sec:2.c}

The detector relies on a common sampling scheme, where the receiver makes $M$ observations in every bit interval. 
The value of the $m$th observation in the $j$th bit interval is labeled $s_{j,m}$. We define the sampling times within a single interval as the function $g(m)$, and the global time sampling function $t(j,m)=jT_{\text{int}}+g(m)$, where $j=\{1,2, \ldots, B\}$ and $m=\{1,2, \ldots, M\}$. In this work, we set $g(m)=m t_s$, implying that the observations are taken at times separated by constant period $t_s$. In addition, we assume that the transmitter and receiver are perfectly synchronized; that is, the transmitter knows the exact time that it has to inject the molecules into the channel, and the receiver detector knows the exact time that it has to sample the number of molecules.

We use the weighted sum detector proposed in \cite{14A.N3} whose decision rule in the $j$th bit interval can be described as
\begin{equation} \label{eq:9}
\hat{W}[j]= 
\left\{\begin{array}{ll}
1& \text{if}~ \sum_{m=1}^M \omega_m N_{A_{obs}}(t(j,m)) \geq \gamma, \\
0& \text{otherwise},
\end{array}\right. 
\end{equation}
where $\omega_m$ is the weight of the $m$th observation and $\gamma$ is the binary decision threshold. In this work, we set $\omega_m=\overline{N_{A_{obs}}}(g(m))$. In~\cite{14A.N3}, it is analytically verified that the matched filter of setting the sample weight $\omega_m$ equal to the mean number of observed molecules $\overline{N_{A_{obs}}}(g(m))$ in \eqref{eq:8} is optimal in the sense that it maximizes the signal-to-noise ratio and minimizes the bit error rate (BER) if the desired signal is corrupted by additive white Gaussian noise. However, channel noise in this work is characterised with a Poisson random variable, which means that the matched filter may not be optimal. Even so, the simplicity of the weighted sum detector, compared to the optimal maximum likelihood detector~\cite{07P.J}, is desirable for practicalities of implementation and given that individual transceivers in MC systems generally have limited computational abilities and memory. In addition, the optimal $\gamma$ can be acquired via numerical search~\cite{14A.N3}.

\section{Time-Varying Electric Fields}
\label{sec:3}

\begin{figure}[t]
    \centering\subfloat[Sinusoidal electric field with $\phi_v=5.07$ (rad).]{
    \includegraphics[width=0.45\textwidth]{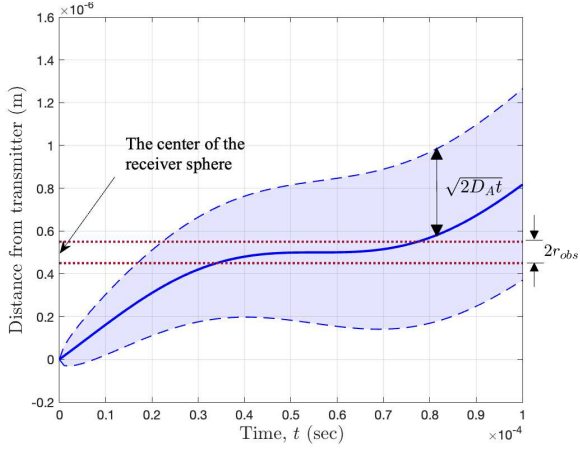}} ~
    \centering\subfloat[Sinusoidal electric field with $\phi_v=\pi$ (rad).]{
    \includegraphics[width=0.45\textwidth]{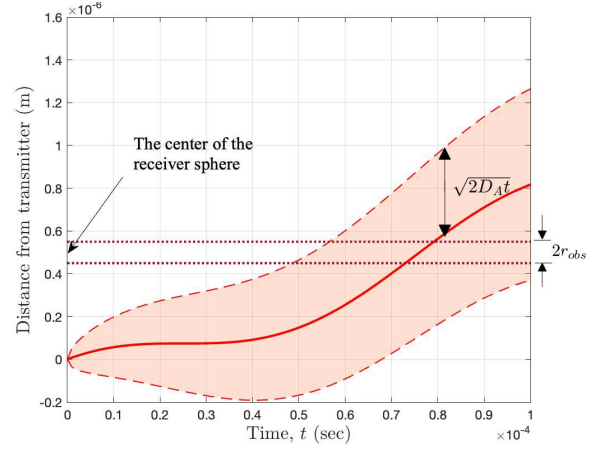}} \\
    \centering\subfloat[Constant electric field, with $v_x(t)=0.01$ (m/s).]{
    \includegraphics[width=0.45\textwidth]{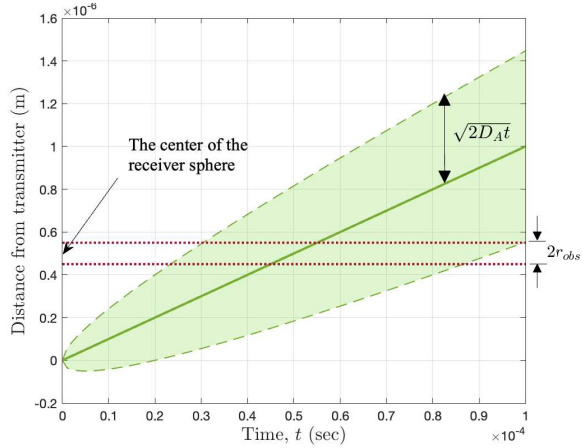}} 
    \caption{The temporal position of the center of the molecule group, resulting from two sinusoidal electric fields (with the different phases $\phi_v$) and a constant electric field benchmark. The sinusoidal velocity component is given as $v_x(t)=8.17\times 10^{-3}\times\sin(2\pi \times 10^4 t - \phi_v) + 8.17\times 10^{-3}$ {(m/s)}, where $\phi_v=5.07$ and $\pi$ (rad) in panels (a) and (b), respectively, and $v_x(t)=0.01$ (m/s) for the constant field in panel (c).}
    \label{fig:2}
\end{figure}

This section investigates two different time-varying (sinusoidal and optimized) electric fields that aim to improve BER performance in MC systems. Bit decoding errors can generally be reduced by increasing the signal strength (i.e., the number of observed molecules at the receiver site) and reducing the ISI between bit intervals (i.e., the number of residual molecules). We propose methods exploiting those time-varying electric fields to fulfill these necessities.

\subsection{Sinusoidal Field}
\label{sec:3.A}

In this section, we consider a sinusoidal electric field, which can be accurately generated and controlled even in small-sized and low-powered lab-on-a-chip applications~\cite{Nyholm2005}. Since $\vec{v}(t) \propto \vec{E}(t)$, we can express the induced sinusoidal molecule velocity (in m/s) as
\begin{equation} \label{eq:sine}
    \vec{v}(t)=\bigl\{A_v \sin(2\pi f_v t - \phi_v) + \text{DC}_v, 0, 0 \bigr\}.
\end{equation}
Without loss of generality, we assume that only $x$-axis flow exists, since the transmitter and receiver are aligned with the $x$-axis.

According to \eqref{eq:2}, the degree of dispersion of the molecules depends on the diffusion coefficient $D_A$ and the time elapsed from being emitted by the transmitter, $t-t_0$. Since $D_A$ is constant for a given fluid medium, we can reduce the degree of molecule dispersion by reducing the time elapsed from emission. But since the expected number of observed molecules $\overline{N_{A_{obs}}}(t)$ is inversely proportional to the degree of the molecules' dispersion at the receiver site, it would be wise to induce a time-varying electric field that sends the molecules toward the receiver as fast as possible immediately after they are emitted. This electric field strategy would allow the molecules to reach the receiver sphere with a higher density. Besides, to allow the detection mechanism described in Section~\ref{sec:2.c} to sample as many molecules as possible (when a binary 1 is transmitted), the time-varying electric field must ensure that the molecules dwell inside the receiver sphere as long as possible once they arrive there. The time-varying electric field should then migrate the molecules away from the receiver sphere just before the next bit interval to reduce the ISI.

Based on this physical mechanism, we provide a method to find rational parameters for the sinusoidal field in terms of the induced sinusoidal velocity \eqref{eq:sine}, i.e. $A_v$, $DC_v$, $f_v$, and $\phi_v$. First, setting $f_v=1/T_{int}$ is a natural choice to have a single velocity fluctuation in a single bit interval since we need to move the molecules twice quickly (towards and away from the receiver sphere) and once slowly (within the receiver sphere). Also, given the average power constraint on the molecule velocity
\begin{equation} \label{eq:10}
\displaystyle{\frac{1}{T_{\text{int}}}\int_0^{T_{\text{int}}} \left(A_v \sin\left(2\pi (1/T_{\text{int}}) t- \phi_v \right) + \text{DC}_v\right)^2 \diff t} \leq \xi_v
\end{equation}
where $\xi_v$ is the constraint value\footnote{Note that this constraint is identical to constrain the average power of electric field since $\vec{v}(t)\propto \vec{E}(t)$.}, it would be necessary to maximize $A_v$ to let the sinusoidal velocity retain the largest gap between its maximum and minimum peak values. This setup enables the molecule group to be quickly sent towards the receiver and ensures that its center stays within the receiver sphere for a long time. Since destructive flows, defined as a flow component not in the direction of transmission (i.e., a negative velocity), generally reduce the peak number of molecules expected to be observed at the receiver~\cite{14A.N}, we constrain the non-zero sinusoidal velocity component to be positive. Thus, by setting $A_v=\text{DC}_v$, $A_v$ and $\text{DC}_v$ can be calculated from \eqref{eq:10} as
\begin{equation} \label{eq:11}
A_v=\text{DC}_v=\sqrt{\frac{2}{3}\xi_v}.
\end{equation}
Regarding $\phi_v$, in order to make the center of the molecule group remain within the receiver sphere for as long as possible, the time that the induced velocity has a minimum should coincide with the time that the center of the molecule group reaches the center of the receiver sphere. This criterion can be expressed as
\begin{equation} \label{eq:12}
\int_{0}^{t_1} \left(A_v \sin\left(2\pi f_v t- \phi_v \right) + \text{DC}_v\right) \diff t = x_0.
\end{equation}
$t_1$ can be numerically found from
\begin{align}
	&v_x'(t_1) = 2 \pi A_v f_v \cos \left(2 \pi f_c t_1 - \phi_v \right) = 0 \nonumber \\
	&\rightarrow t_1 = \frac{1}{ 2 \pi f_v} \left( \frac{\pi}{2} (2n-1)+ \phi_v \right)~\text{for}~n\in \mathbbm{Z}_+ \label{eq:13}
\end{align}
where $\mathbbm{Z}_+$ is the set of non-negative integers. Then, $\phi_v$ can be numerically found from \eqref{eq:12}. Note that the method explained above is not proved optimal. However, as will be shown in Section \ref{sec:5}, the induced sinusoidal velocity obtained via this method significantly increases the expected number of observed molecules, compared to the constant electric field benchmark.

Fig.~\ref{fig:2} illustrates how the time-varying sinusoidal electric field moves the molecules throughout a bit interval. The center of the molecule group emitted at $t=0$ is shown according to the time elapsed, both when a sinusoidal electric field (two phases, $\phi_v=5.07$ and $\pi$ (rad)) and a constant electric field are applied. The values on the vertical axis of the figures are calculated as $\int_{0}^{t} v_x(\tau) \diff \tau$. For the sinusoidal electric field in Fig.~\ref{fig:2}(a), $v_x(t)=8.17\times 10^{-3}\sin(2\pi \times 10^4 t - 5.07) + 8.17\times 10^{-3}$ (m/s) is obtained from the method explained above. In contrast, the second sinusoidal induced velocity $v_x(t)=8.17\times 10^{-3}\sin(2\pi \times 10^4 t - \pi) + 8.17\times 10^{-3}$ (m/s) and the constant induced velocity $v_x(t)=0.01$ (m/s) in Figs.~\ref{fig:2}(b) and (c), respectively, are illustrative examples for comparison. All of the velocities have the same average power as $10^{-4}$. In addition, the standard deviation of the molecules' distribution from the center of the molecule group, $\sigma_x=\sqrt{2 D_A t}$, is shaded. Note that the transmitter and the receiver sphere are located at $\{-x_0, 0, 0\}$ and $\{0, 0, 0\}$, respectively; thus, the distance between the transmitter and receiver is $x_0$ ($=\SI{0.5}{\micro\meter}$ in the figure). The receiver sphere region (spanning $2r_{obs}$ along the $x$-axis) is delineated by the two dashed horizontal lines. All the other system parameters are given in Table~\ref{table:1} (Section~\ref{sec:5}).

In Fig.~\ref{fig:2}(a), the center of the molecule group (the solid blue line) quickly reaches the receiver sphere and remains within the receiver sphere region for a long time. In contrast, in Fig. \ref{fig:2}(b) and (c) the molecule group slowly reaches, and quickly passes through, the receiver sphere. Moreover, the degree of molecule dispersion in Fig.~\ref{fig:2}(a) when the center of the molecule group reaches the receiver sphere is smaller than that in either Fig.~\ref{fig:2}(b) or (c). Note that the molecules' standard deviation from the center of the molecule group is proportional to the elapsed time $t$. Since the density of molecules at the receiver site in Fig.~\ref{fig:2}(a) is higher, the expected number of observed molecules $\overline{N_{A_{obs}}}(t)$ when the center of the molecule group stays in the receiver sphere is also higher than those of Fig.~\ref{fig:2}(b) and (c). Besides, for the sampling scheme described in Section~\ref{sec:2.c}, which makes multiple observations at times separated by a constant interval, an increase in the duration that the molecules dwell in the receiver sphere increases the value of the weighted sum in \eqref{eq:9} (when a binary 1 is transmitted). These features of the sinusoidal electric field with phase $\phi_v = 5.07$, shown in Fig.~\ref{fig:2}(a), significantly improve the bit error performance compared to the two fields considered. This statement will be verified in Section~\ref{sec:5}.

\subsection{Optimized Field}
\label{sec:3.B}

Further developing the advantages of the sinusoidal electric field discussed in Sec.~\ref{sec:3.A}, this section analytically derives an optimized electric field that colocates the center of the molecule group and the receiver sphere for as long as possible. First, we formulate a cost function of the root-mean-square error between the receiver's center and the center of the molecule group as \begin{subequations}\label{eq:14}
\begin{align} \label{eq:14a}
&C[x, x'] = \displaystyle\sqrt{\frac{1}{T_{\text{int}}} \int_{0}^{T_{\text{int}}} \left(x(t) - x_0 \right)^2 \diff t} \\
&\text{s.\,t.}\, \left\{ \begin{array}{l}
\displaystyle{\frac{1}{T_{\text{int}}}\int_0^{T_{\text{int}}} v_x^2(t) dt} \leq \xi_v, \\
IC: x(0)=0, \\
FC: x(T_{\text{int}})=x_1, \end{array} \right. \label{eq:14b}
\end{align}
\end{subequations}
where $x(t)=\int_{0}^{t} v_x(\tau) \diff \tau$. 
The first constraint limits the average power of electric field, and the final condition $FC$ is set in order to migrate the molecules in the current bit interval away for the new molecules coming in the next bit interval. Then, letting $x'(t)=v_x(t)$, we form the Lagrangian
\begin{equation} \label{eq:15}
\mathcal{L}[x(t), x'(t)]=\left[x(t)-x_0\right]^2 + \mu \left[ x'(t) \right]^2,
\end{equation}
where $\mu \in \mathbb{R}$ is the Lagrange multipliers. Therefore, the Euler-Lagrange equation~\cite{14D.B} can be written as
\begin{equation} \label{eq:16}
\frac{\diff}{\diff t}\left( \frac{\partial \mathcal{L}}{\partial x'(t)}  \right) - \frac{\partial \mathcal{L}}{\partial x(t)}=\frac{\diff}{\diff t} \left(2 \mu x'(t) \right) - 2\left(x(t)-x_0\right) = 0.
\end{equation} 
Letting $y(t)=x(t)-x_0$, so $y'(t)=x'(t)$, we can rewrite \eqref{eq:16} as
\begin{equation}\label{eq:17}
y''(t) - \lambda^2 y(t) = 0,
\end{equation}
where $\mu=1/\lambda^2$. Therefore, by utilizing the initial and final conditions ($IC$ and $FC$) in \eqref{eq:14b}, we can obtain the solution of $x(t)$ as
\begin{equation} \label{eq:18}
x^*(t)=C_1 e^{\lambda t}+C_2 e^{-\lambda t} + x_0,
\end{equation}
where 
\begin{subequations}
\begin{align} \label{eq:19}
&C_1=\frac{(1-e^{-T_{\text{int}}\lambda})x_0-x_1}{e^{-T_{\text{int}}\lambda}-e^{T_{\text{int}}\lambda}},  \\
&C_2=-x_0-\frac{(1-e^{-T_{\text{int}}\lambda})x_0-x_1}{e^{-T_{\text{int}}\lambda}-e^{T_{\text{int}}\lambda}}.
\end{align}
\end{subequations}
$\lambda$ can be numerically found from the first constraint in \eqref{eq:14b} as
\begin{equation} \label{eq:20}
\frac{\lambda}{2T_{\text{int}}} \left[ C_1^2 (e^{2 T_{\text{int}}\lambda}-1) + C_2^2(-e^{-2 T_{\text{int}}\lambda}+1) -4C_1 C_2 T_{\text{int}}\lambda \right]=\xi_v.
\end{equation}
Finally, the velocity induced by the optimized electric field can be obtained as 
\begin{equation} \label{eq:exponential}
    v_x^*(t)=(x^*(t))'=C_1 \lambda e^{\lambda t} - C_2 \lambda e^{-\lambda t}~~\text{(m/s)},
\end{equation}
which has a form of exponential functions.

\begin{figure}[t]
    \centering\subfloat[$x(t)$]{
    \includegraphics[width=0.51\textwidth]{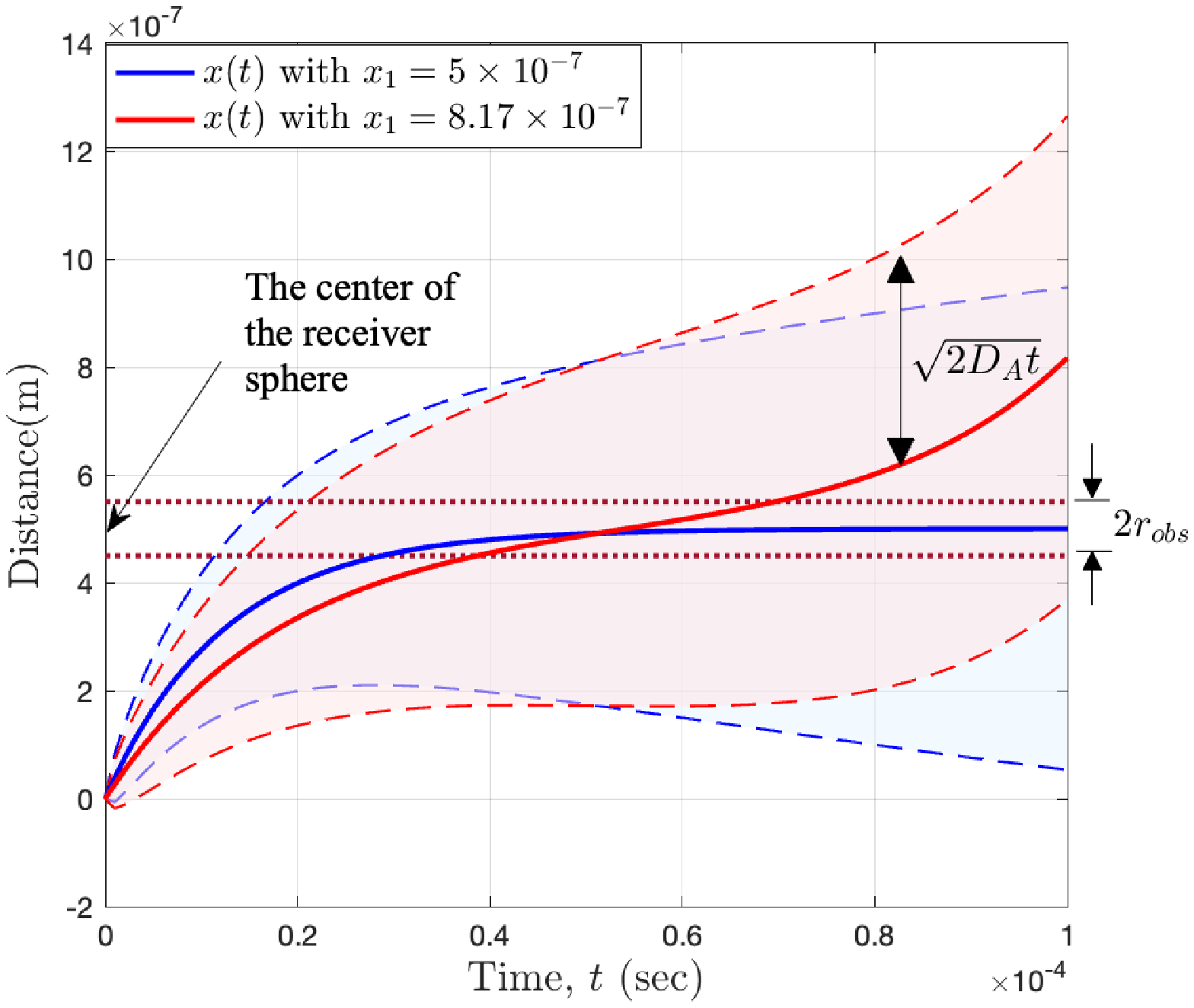}} 
    \centering\subfloat[$v_x(t)$]{
    \includegraphics[width=0.51\textwidth]{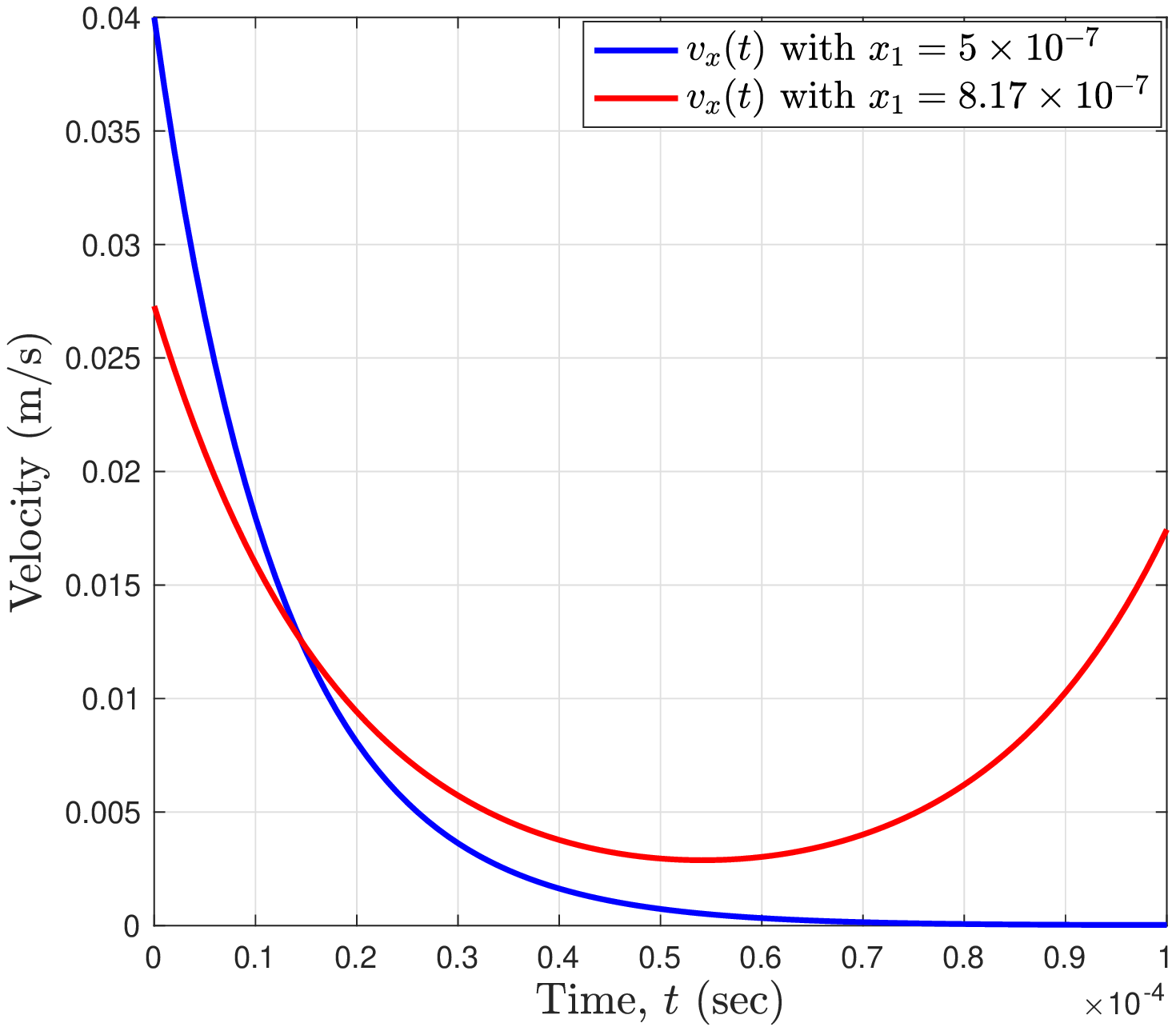}} 
    \caption{The location of the center of the molecule group and the $x$-axis velocity $v_x(t)$ of the molecules according to the elapsed time, when the optimized electric field is applied (with velocity obtained from \eqref{eq:14}). $\xi_v=10^{-4}$ is assumed.}
    \label{fig:3}
\end{figure}

Fig~\ref{fig:3} illustrates the location of the center of the molecule group and the $x$-axis velocity of the molecule flow according to the elapsed time when the optimized electric field \eqref{eq:14} is employed with the different final conditions $x_1$. The bit time interval $T_{int}=10^{-4}$ (sec) is used. Note that the final condition $x_1=8.17\times10^{-7}$ (m) is identical to the travel distance of the center of the molecule group in a single bit interval time with the sinusoidal velocity in Figs.~\ref{fig:2}(a) and (b). From the figures, we can note that the obtained optimized electric fields initially propagates the molecules at high speed immediately after emission, then rapidly reduce their speed to ensure that the molecules remain at the receiver sphere for a long time. Just before the next bit interval, the velocity is increased to migrate the molecules away to mitigate ISI (when the final condition is $x_1>x_0$).

\section{Balance Between Electrophoretic and Viscous Forces}
\label{sec:4}

In previous sections, we assumed that the electrophoretic force $\vec{F}_E(t)=q_A\vec{E}(t)$ induces molecular motion with velocity $\vec{v}(t)$ via a linear relation, $\vec{E}(t) \propto \vec{v}(t)$. However, the \emph{instantaneous} molecule velocity $\vec{u}(t)$ may not necessarily equal $\vec{v}(t)$ due to a viscous drag force and an added mass effect, especially during the period of initial acceleration following emission that is implied by both the sinusoidal and optimized fields from Section~\ref{sec:3}. Therefore, in this section, we derive and analytically solve an ordinary differential equation (ODE) for the time-varying sinusoidal and exponential electric fields to determine when $\vec{u}(t) \approx \vec{v}(t)$ for $t \neq 0$ is valid.

In fluid dynamics, the Basset–Boussinesq–Oseen (BBO) ODE describes the motion of -- and forces on -- a small particle/molecule in unsteady flow at low Reynolds number \cite{SommerfeldTB}. This equation considers 1) the viscous drag, 2) the added mass effect, 3) the pressure gradient force due to an unsteady undisturbed flow, 4) the Basset force (a history force due to the non-instantaneous molecule boundary layer development), and 5) other body forces, such as the electrophoretic force and gravity. First, the viscous drag for a spherical molecule is given by Stokes' law as $F_D(t)=f \vec{u}(t)$ in the limit of a small Reynolds number, where $f=6 \pi \mu_f r_m$ is the frictional drag coefficient, $\mu_f$ is the dynamic viscosity of the fluid, and $r_m$ is the molecule radius. Second, the added mass effect arises due to the displacement of fluid required to accelerate the molecule through the ambient fluid~\cite{SommerfeldTB}. For a spherical molecule, the added mass comprises a fluid region of half the molecule volume. Moreover, we assume an undisturbed flow (i.e. zero pressure gradient force) and neglect gravity due to the small molecule size. We also neglect the Basset force, which is a common assumption in the literature for conceptual simplicity and flow considerations~\cite{EslamiPhysFluids2016,RuijinIJHMT2019}. The assumptions made here are reviewed in Section~\ref{sec:6}.

Based on these assumptions, we formulate a first-order BBO ODE as
\begin{equation}
    \label{eq:BBO}
    \frac{2}{3}\pi r_m^3 (2\rho_m + \rho_f)\frac{\diff \vec{u}(t)}{\diff t} + 6 \pi \mu_f r_m\vec{u}(t) = q_A \vec{E}(t),
\end{equation}
where $\rho_m$ and $\rho_f$ are the molecule and fluid densities, respectively \cite{EslamiPhysFluids2016}. We consider a \emph{single} molecule introduced at $t=0$ with zero velocity; thus, the initial condition is $\vec{u}(0) = \vec{0}$. Building on the given assumption that the electric field $\vec{E}(t)$ is linearly related to the molecule velocity $\vec{v}(t)$, we set $\vec{E}(t) = (f/q_A)\vec{v}(t)$ and plug this into \eqref{eq:BBO}. Here, $\vec{v}(t)$ is interpreted as the \emph{desired} velocity of the molecule (that we want the molecule to achieve), and $\vec{E}(t)$ is the \emph{applied} electric field required to induce the desired velocity $\vec{v}(t)$. In other words, by solving the ODE \eqref{eq:BBO} for $\vec{u}(t)$, we can investigate the difference between the instantaneous molecule velocity $\vec{u}(t)$ induced when $\vec{E}(t)$ is applied, and the desired molecule velocity $\vec{v}(t)$.

Note that the left side of \eqref{eq:BBO} can be divided into mass (first) and viscous (second) terms. To make $\vec{u}(t) \approx \vec{v}(t)$, the viscous term must dominate the mass term throughout the dynamics when $\vec{v}$ is not close to zero. This condition can be expressed as
\begin{equation}
    \frac{2}{3}\pi r_m^3 (2\rho_m + \rho_f) \left\vert \frac{\diff \vec{u}(t)}{\diff t} \right\vert \ll 6 \pi \mu_f r_m \vert \vec{u}(t)\vert,
\end{equation}
which provides the feasible region for the particle radius $r_m$ as
\begin{equation}
    \label{eq:constraint}
     r_m \ll \sqrt{\frac{9\mu_f \vert\vec{u}(t)\vert}{(2\rho_m + \rho_f) \left\vert\diff \vec{u}(t)\middle/\diff t\right\vert} }.
\end{equation}

\begin{figure}[t]
    \centering\subfloat[$v_x = 8.17\times 10^{-3}\sin(2\pi \times 10^4 t - 5.07) + 8.17\times 10^{-3}$~(m/s)]{
    \includegraphics[width=0.5\textwidth]{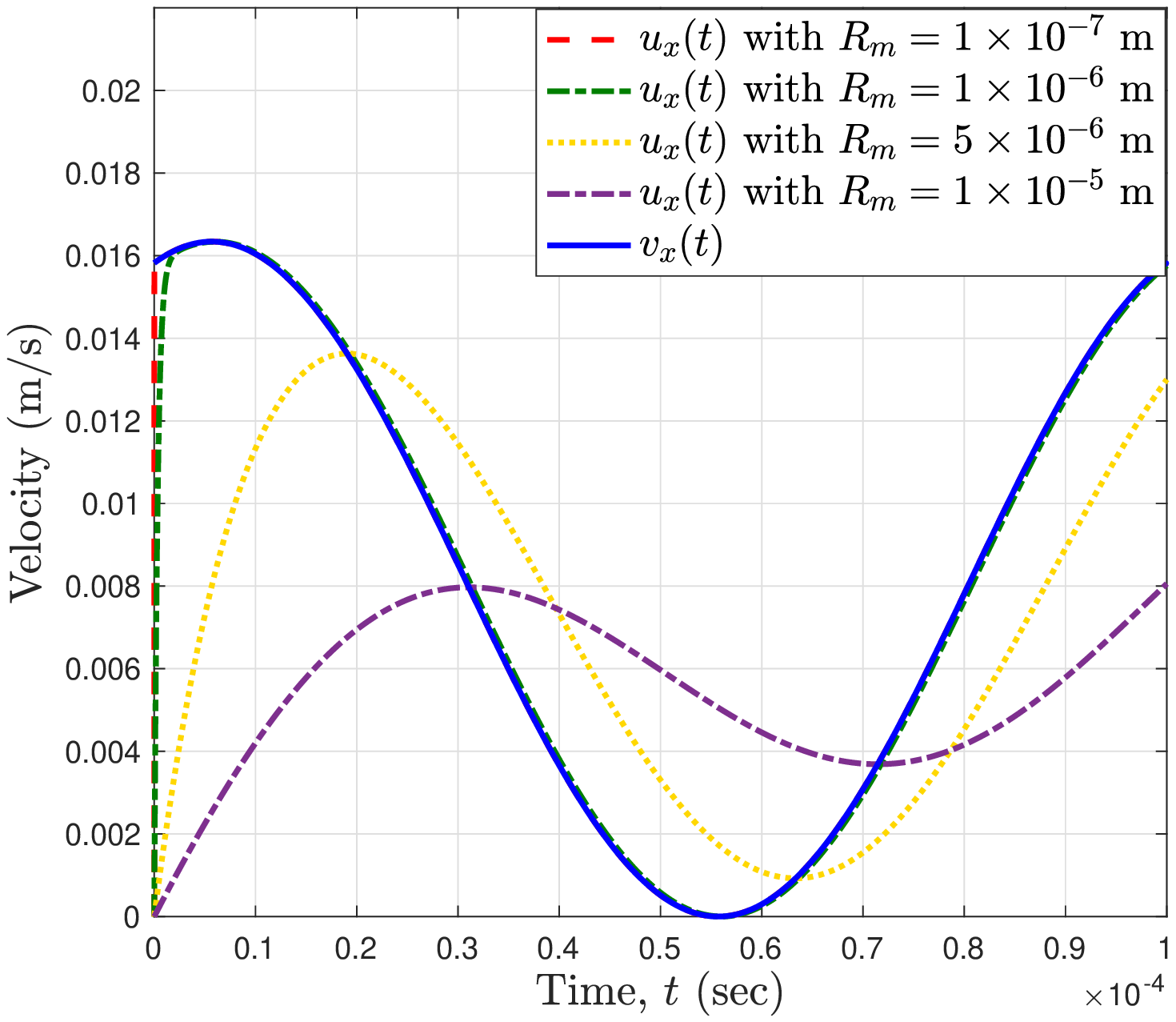}} ~
    \centering\subfloat[$v_x = 0.04 e^{-80000 t}$~(m/s)]{
    \includegraphics[width=0.5\textwidth]{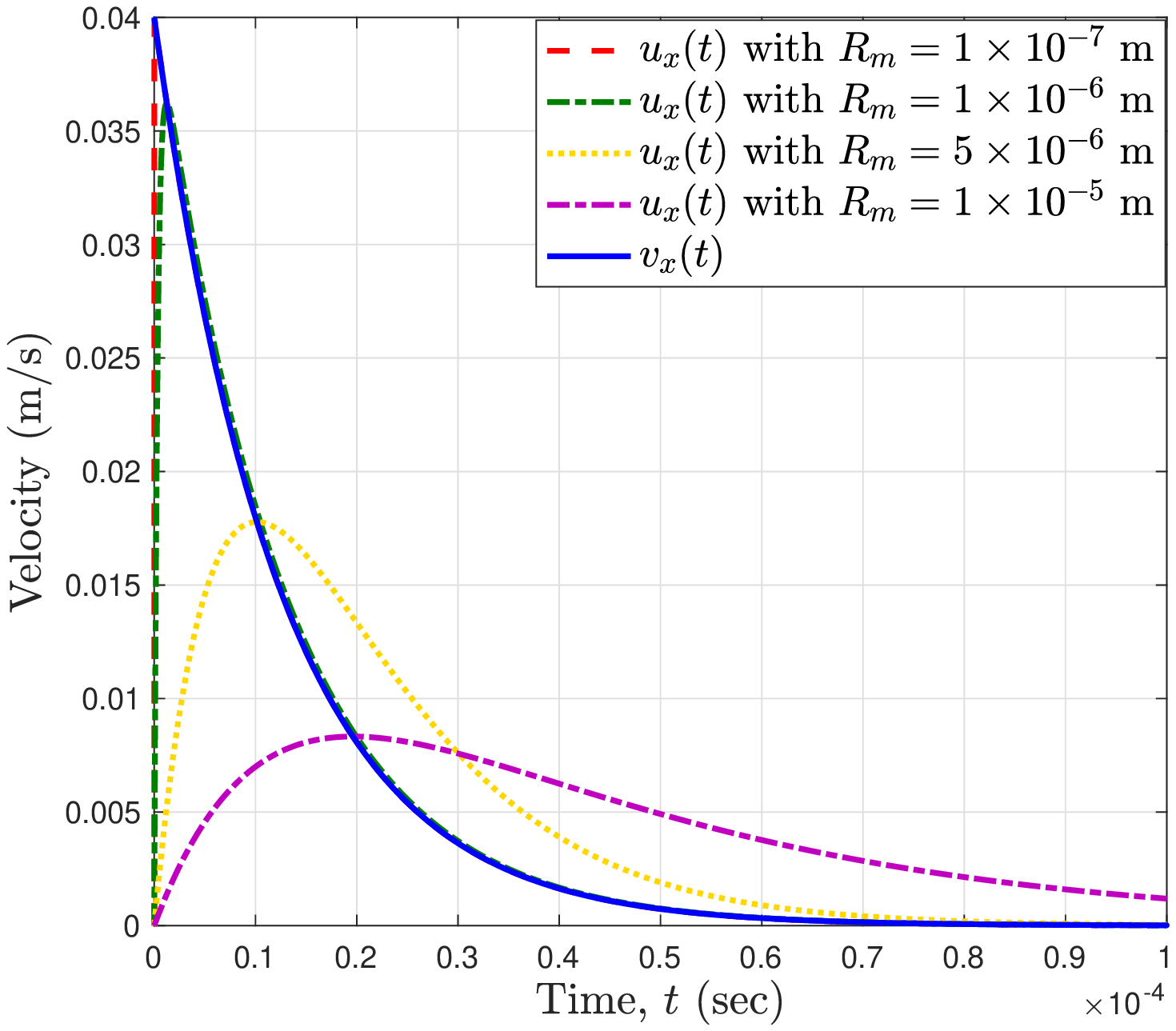}} 
    \caption{Comparison between instantaneous velocities $u_x(t)$ (accounting for mass and viscosity) for different molecule radii and the desired velocity $v_x(t)$. The sinusoidal and optimized fields from Fig.~\ref{fig:2}(a) and Fig.~\ref{fig:3}(b), respectively, are assumed.}
    \label{fig:4}
\end{figure}

We now consider two examples based on the time-varying electric fields studied in Section~\ref{sec:3} to elucidate \eqref{eq:constraint}. We assume flow only in the $x$ direction, so only that vector component appears here. First, for the sinusoidal field, \eqref{eq:BBO} can be integrated (using \eqref{eq:sine}) to
\begin{align}
  u_x(t) =& \frac{f\Omega}{\Gamma}\sin\bigl( 2\pi f_vt - \phi_v \bigr) - 2\pi f_v\Omega\cos\bigl( 2\pi f_vt - \phi_v \bigr) + DC_v \nonumber\\
  &+ \left[ \frac{f\Omega}{\Gamma}\sin\bigl(\phi_v\bigr) + 2\pi f_v\Omega\cos\bigl(\phi_v\bigr) - DC_v \right] \exp\left( -\frac{ft}{\Gamma} \right),
\end{align}
where
$$ \Omega = \frac{\Gamma A_vf}{ f^2 + \bigl( 2\pi f_v\Gamma\bigr)^2}  ~~\text{and}~~ \Gamma = \frac{2}{3}\pi r_m^3\bigl(2\rho_m + \rho_f\bigr).$$
Employing the sinusoidal velocity parameters as used in Fig.~\ref{fig:2}(a), and the environment parameters $\rho_{m}=\rho_{f} = 10^3~\si{\kilo\gram\per\meter\cubed}$ and $\mu_f = 10^{-3}~\si{\pascal\second}$, $u_x(t)$ is plotted over a single bit interval time for different values of $r_m$ in Fig.~\ref{fig:4}(a). $v_x(t)$ is also plotted for comparison. Second, for the optimized field introduced in Sec.~\ref{sec:3.B} with the final condition $x_1=x_0$ (i.e., $C_1=0$), \eqref{eq:BBO} can be integrated to obtain
\begin{equation}
    \label{eq:expSol}
    u_x(t) = \frac{\lambda f C_2}{f - \lambda \Gamma} \left[ \exp\bigl( -ft/\Gamma \bigr) - \exp\bigl( -\lambda t \bigr) \right],
\end{equation}    
with $v_x(t)$ given by \eqref{eq:exponential}. Using the parameters from Fig.~\ref{fig:3}(b), namely $C_2 = -\num{5e-7}$ and $\lambda = \num{8e4}$, $u_x(t)$ is plotted for different values of $r_m$, alongside $v_x(t)$, in Fig.~\ref{fig:4}(b).

For both cases in Figs.~\ref{fig:4}(a) and (b), $u_x(t)$ for $r_m \lessapprox \SI{1e-6}{\meter}$ shows insignificant deviation from $v_x(t)$, verifying that the applied electric field can linearly induce the molecule velocity $\vec{v}$ for such small molecule sizes. For $r_m = \SI{1e-6}{\meter}$, although there is a degree of deviation between $u_x(t)$ and $v_x(t)$ during the initial acceleration of the former, the instantaneous velocity quickly grows enough that the deviation becomes negligible after only a few microseconds. However, for larger molecule radii (notably $r_m = \SI{5e-6}{\meter}$ and $\SI{1e-5}{\meter}$), a persistent lag between $u_x(t)$ and $v_x(t)$ exists, which shows that the electric field cannot linearly produce the desired molecule velocity when the molecule radius is not small.

The main implication of Fig.~\ref{fig:4} is that for $\vec{u}(t) \approx \vec{v}(t)$, $\vec{u}$ (i.e. the viscous term) must grow sufficiently quickly from $t = 0$ to balance $\vec{v}$, where the latter is not necessarily zero at $t = 0$. We can relate this requirement to \eqref{eq:constraint}, which thus needs to be first satisfied shortly after $t = 0$. For example, with the sinusoidal electric field ($\phi_v = 5.07$) considered above, \eqref{eq:constraint} is first satisfied (taken here to mean that the right-hand side is at least $10r_m$) within 1.7\% of the bit interval when $r_m = \SI{1e-6}{\meter}$, whereas 18.7\% of the bit interval is required for $r_m = \SI{5e-6}{\meter}$. Hence, in the former case the viscous term quickly dominates the mass term after $t=0$, but not in the latter case, which explains the deviation seen between $u_x(t)$ and $v_x(t)$ in Fig.~\ref{fig:4}(a) when $r_m = \SI{5e-6}{\meter}$ from a physical perspective. Similar conclusions can be drawn for the optimized electric field displayed in Fig.~\ref{fig:4}(b).

For the system model considered in this work (see Table~\ref{table:1}), the diffusion coefficient $D_A = 10^{-9}~\si{\meter\squared\per\second}$ implies a molecule radius of $r_m = \SI{2.2e-10}{\meter}$ with a room temperature environment, via the Stokes-Einstein equation. Based on the above analysis, this implied radius is several orders of magnitudes lower than those for which the viscous term looses its dominance, demonstrating that electrophoretic MC is feasible from a fluid dynamics perspective. Hence, under the conditions of typical electrophoretic MC systems, and in particular throughout all other sections of this work, we can reliably say that $\vec{u}(t) \approx \vec{v}(t)$ for $t \neq 0$ and assume equivalency between the magnitudes of $\vec{E}(t)$ and $\vec{v}(t)$.

\section{Numerical results}
\label{sec:5}

\begin{table} [t]
\centering 
\caption{System parameters} \small
\begin{tabular}{ l | c | c }
   \Xhline{2\arrayrulewidth}
  Parameter & Symbol & Value \\ \Xhline{2\arrayrulewidth}
  Number of molecules per emission & $N_{EM}$ & $10^4$   \\  \hline
  Probability of binary 1 & $P_1$ & 0.5  \\  \hline
  Length of transmitter sequence & $B$ & 100 bits  \\  \hline
  Bit interval time & $T_{\text{int}}$ & \SI{0.1}{\milli\second}  \\  \hline
  Diffusion coefficient & $D_A$ & $10^{-9}~\si{\meter\squared\per\second}$ \\  \hline
  Location of transmitter & $x_0$ & \SI{0.5}{\micro\meter} \\  \hline
  Radius of receiver & $r_{obs}$ & \SI{50}{\nano\meter} \\  \hline
  Expected impact of noise source(s) & $\overline{N_{A_{n}}}(t)$ & 1 molecule \\  \hline
  The number of samples at detector & $M$ & 5 \\  \hline
  The average power constraint & $\xi_v $ & $10^{-4}$ \\ \hline
  Simulation step size & $\Delta t$ & \SI{0.5}{\micro\second} \\  \Xhline{2\arrayrulewidth}
\end{tabular} \label{table:1}
\end{table}

In this section, we provide numerical results to verify our analysis from the section above. The system parameters are provided in Table~\ref{table:1}. The simulation results were obtained via a Monte Carlo approach with $10^4$ trials. In the simulations, we generate Poisson random variables with the time-varying mean $\overline{N_{A_{obs}}}(jT_{\text{int}}+g(m))$ to mimic the observation value $s_{j,m}$, and estimate the transmitted binary sequence $\mathbf{W}$ based on the generated values of $s_{j,m}$ by using the weighted sum detector \eqref{eq:9}. The BER is subsequently calculated. This approach can be justified by the fact that $s_{j,m}$ is approximated well by the Poisson random variables with time-varying mean~\cite{14A.N3}. Note that the independence between adjacent observations becomes negligible as the sampling interval $t_s$ increases. This simulation approach significantly reduces the computational cost compared to a microscopic simulation approach.

\begin{figure}[t]
    \centering
    \centerline{\includegraphics[scale=0.58]{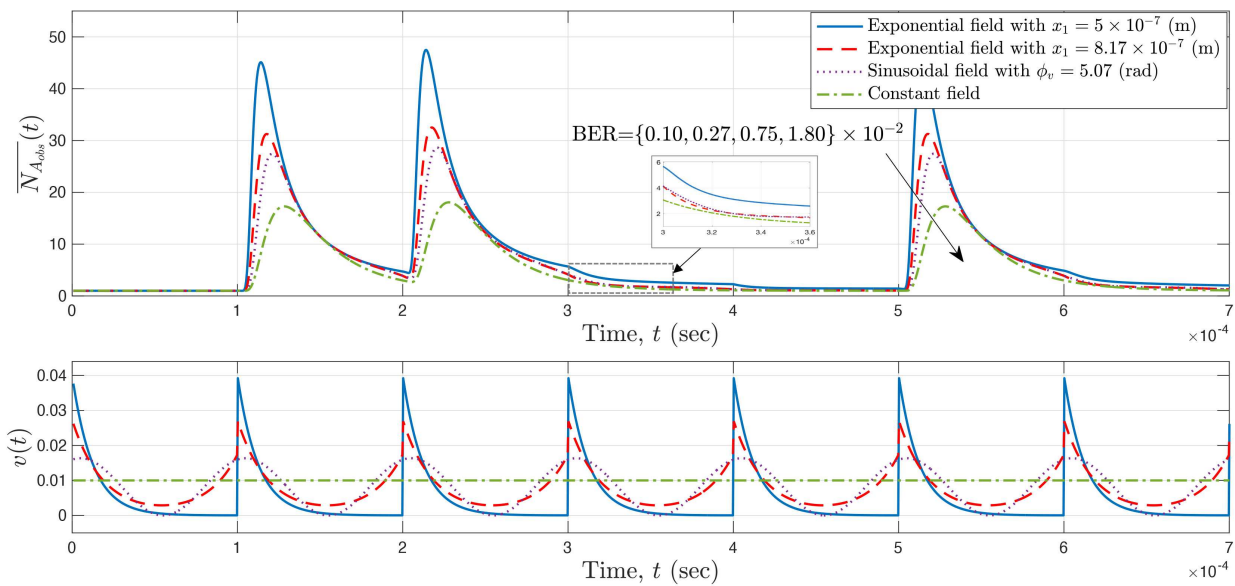}}
    \caption{The expected number of molecules observed within the receiver volume  $\overline{N_{A_{obs}}}(t)$ for the different types of electric field, and therefore molecule velocities $v_x(t)$.}
    \label{fig:5}
\end{figure}

Fig.~\ref{fig:5} shows the expected number of observed molecules within the receiver volume $\overline{N_{A_{obs}}}(t)$ and the $x$-axis velocity component $v_x(t)$ when the different electric fields are employed. The assumed transmitted binary sequences are fixed as $\{0, 1, 1, 0, 0, 1, 0, \ldots \}$ for all the cases, and only the first 7 bits are illustrated in the figures. Note that the optimized electric field, with the final condition $x_1=x_0=5\times10^{-7}$, yields the highest expected number of observed molecules, while the constant field leads to the lowest number. It is also shown that the number of the residual molecules in the next time slot (i.e., the ISI) is also the highest under the optimized electric field (see the region of $0.3 \sim 0.4 \times 10^{-3}$ (sec)), which might adversely affect the BER performance. However, the fact that the BER result of $\{0.10, 0.27, 0.75, 1.80\}\times 10^{-2}$ for each velocity improves (decreases) as the expected number of observed molecules increases verifies that increasing the number of observed molecules (i.e. propagation efficiency) is more important for reducing bit errors than reducing the ISI with the considered detection scheme. It is worth noting that the exponential electric field is optimized with respect to reducing the root-mean-square error in \eqref{eq:14a}, rather than minimizing the ISI; however, we can change the ISI by manipulating the final condition $x_1$ in \eqref{eq:14b} if required.

\begin{figure}[t]
    \centering\subfloat[Bit interval times, $T_{int}$.]{
    \includegraphics[width=0.5\textwidth]{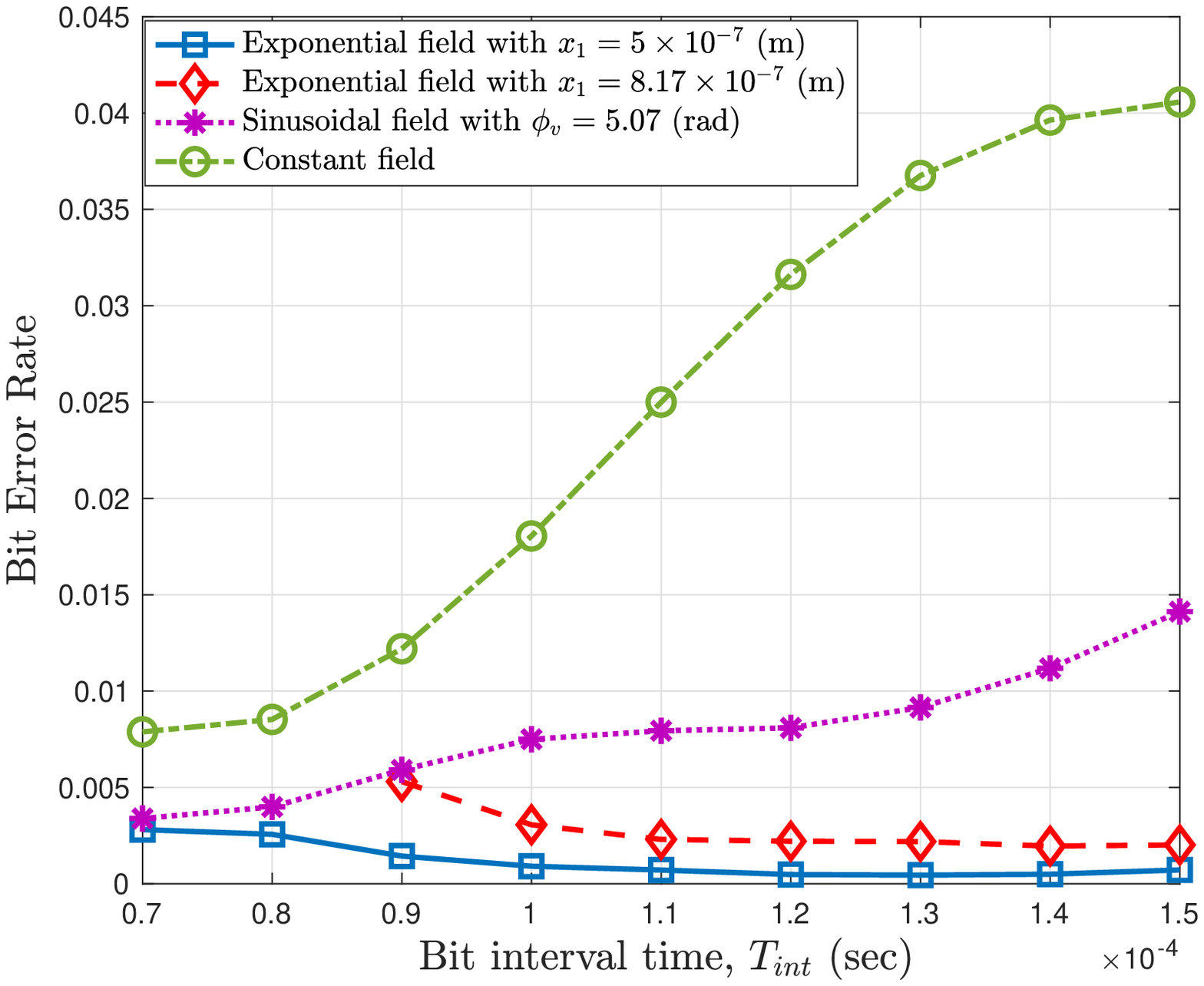}} 
    \centering\subfloat[Average power constraints, $\xi_v$.]{
    \includegraphics[width=0.5\textwidth]{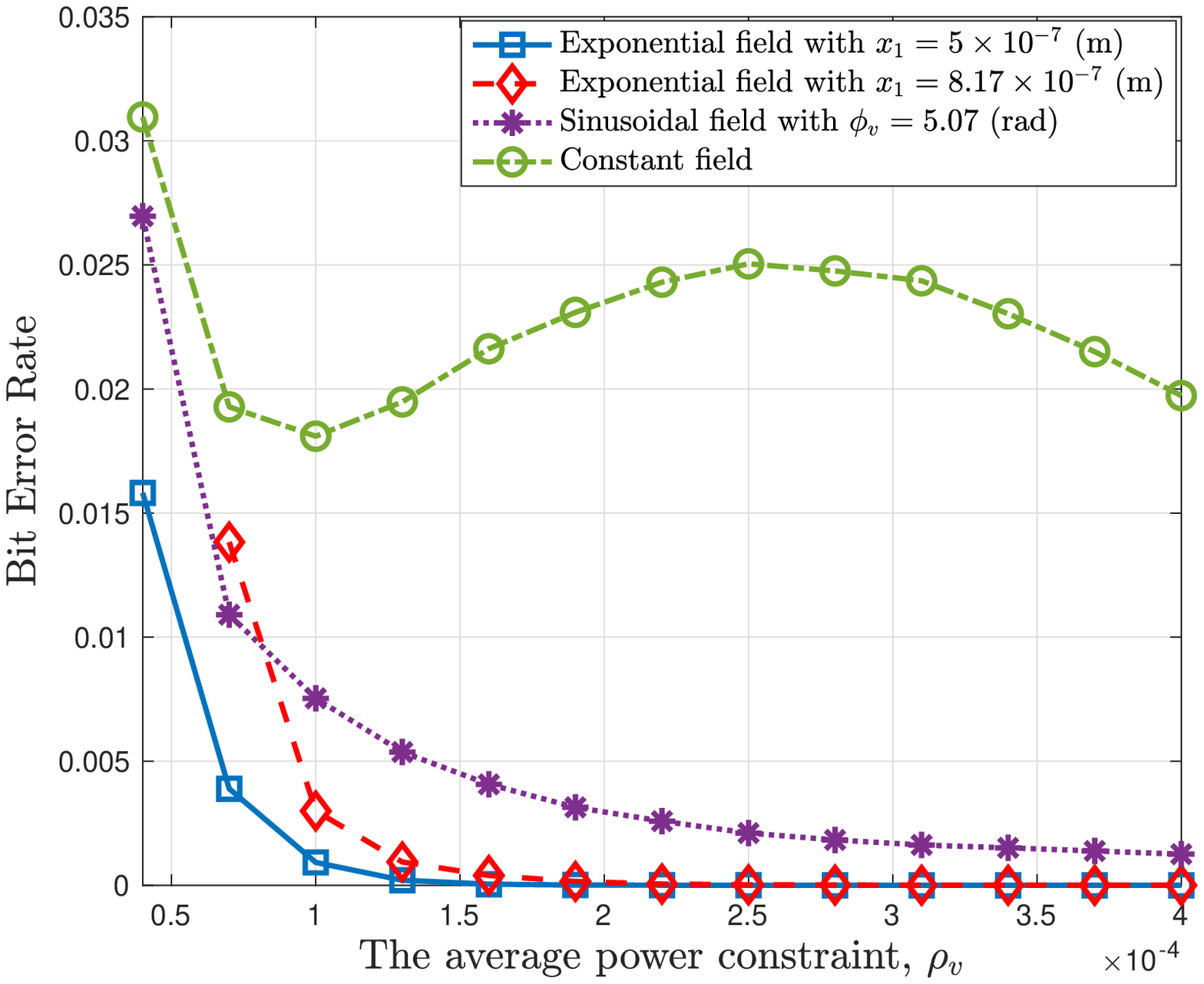}} \\
    \centering\subfloat[Number of samples per a bit interval, $M$.]{
    \includegraphics[width=0.5\textwidth]{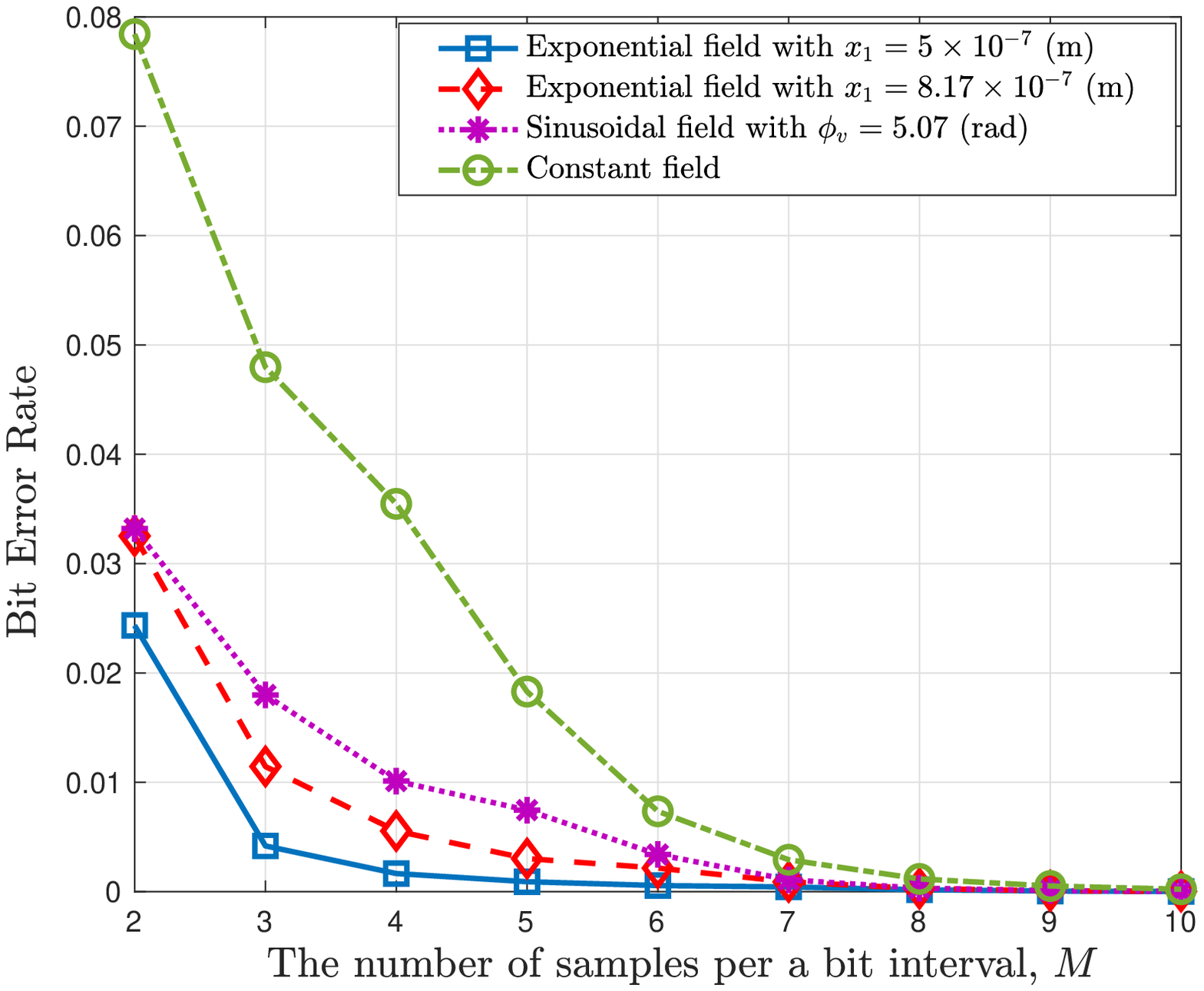}}    
    \caption{BER for different system parameters.}
    \label{fig:6}
\end{figure}

Fig.~\ref{fig:6}(a) shows the BER performance for the different bit interval times $T_{int}$. In conventional wireless or wired communications, when $T_{int}$ increases, it is expected that ISI would decrease so that the BER would also decrease. However, Fig.~\ref{fig:6} demonstrates that the BER increases as $T_{int}$ increases for both the sinusoidal and optimized electric fields. Although ISI between the bit intervals can decrease with increased bit interval time, the molecules can also diffuse more widely. Thus, the expected number of observed molecules over a single bit interval decreases, and the BER consequently worsens. In contrast, since the optimized electric field quickly propagates the molecules and ensures that the center of the molecule group stays within the receiver sphere for a long time, its BER can reduce as $T_{int}$ increases.

Fig.~\ref{fig:6}(b) shows the BER performance for the different average power constraints $\xi_v$. The BER performance curve is not monotonic when the constant field is employed. Since this work employs a constant sampling time $t_s$ as described in Section~\ref{sec:2.c}, the BER performance appears good when the center of the molecule group passes through the receiver sphere at the times that the observations are taken, i.e. $j T_{int} + m \left(T_{int}/M \right)$, where $j=\{1,2, \ldots, B\}$, $m=\{1,2, \ldots, M\}$. In contrast, the BER performance worsens when the crossing time of the center of the molecules group and the receiver sphere deviates from the sampling points. In other words, the BER performance for a constant electric field is dependent on the sampling time, which must be taken into account in engineered MC systems with advection. In contrast, when the sinusoidal and optimized electric fields are employed, it is shown that the BER performance improves as the allowed average power $\xi_v$ increases, independent of the sampling time. This improvement in performance is because these fields propagate the center of the molecule group toward the receiver site more quickly, and make it stay longer at the receiver sphere as $\xi_v$ increases.

{Fig.~\ref{fig:6}(c) shows the BER performance for different numbers of samples per bit interval $M$. As the number of samples increases, the BER performance for all fields improves. This effect arises because the probability that the observations are taken when the centre of the molecule group is close to the center of the receiver sphere rises with larger $M$; thus, the weighted sum in \eqref{eq:9} (when a binary 1 is transmitted) also increases, which results in a lower BER.}

\section{Discussion}
\label{sec:6}

This work was the first trial aiming to improve MC performance by using electrophoresis to induce time-varying molecule velocities and therefore utilized various assumptions of ideal system environments and parameters to reduce the communication model's complexity. These simplifications enabled us to concentrate on the feasibility and advantage of time-varying electric fields to improve MC performance; however, they may also yield various implementation challenges when applying the considered electrophoretic MC in real nanonetwork applications.

For example, we assumed an infinite fluid environment. However, a more realistic channel would be bounded channels, such as circular and rectangular duct channels~\cite{18W.W}. In these bounded channels, the diffusion and flow of the information-carrying molecules would be spatially bounded, significantly influencing the receiver signal model. There may also be specific wall effects. In particular, bounded channels would see an increase the ISI between bit intervals because previously transmitted molecules could not diffuse away to infinity. Therefore, it would be an essential and compelling future work to examine how bounded channels affect communication performance in electrophoretic MC.

Additionally, we assumed a point transmitter with an impulsive injection method that introduces many molecules ($N_{EM}$) at a single point in space and time. It is well known that a molecule's motion is influenced (hindered) by those around it, while the analysis of fluid dynamic feasibility in Section~\ref{sec:4} considered a single isolated molecule. Moreover, electrostatic repulsion between the like-charged molecules will induce additional motion not accounted for here. We hypothesize two scenarios: 1) significant electrostatic repulsion, so the molecules travel individually, and hindrance can be accounted for in \eqref{eq:BBO} with an effective fluid viscosity \cite{antonopoulou2018}; 2) the molecules remain concentrated and travel towards the receiver as a  larger collective (porous) particle. Either case will tighten the molecule radius constraint \eqref{eq:constraint} for which $\vec{u}(t) \approx \vec{v}(t)$, although the assumption is likely still to hold for typical molecule sizes $\mathcal{O}\bigl(10^{-10}\bigr)~\si{\meter}$ of interest in MC applications. Also neglected in \eqref{eq:BBO} is the viscous molecule history effect (Basset force, as noted in Section~\ref{sec:4}), which may play a non-negligible role, especially for the optimized electric field due to its impulse-like manifestation in the molecule velocity \cite{SeylerPhysRevResearch2019}, in addition to any non-sphericity of the molecules. These physical features concerning the point transmitter need to be additionally considered to implement electrophoretic molecular communication with time-varying electric fields in practice.

\section{Conclusions}
\label{sec:7}

We studied electrophoretic molecular communication systems with a time-varying flow of information-carrying molecules to achieve enhanced communication performance. Sinusoidal and exponential electric fields were used to induce molecule flow.  These were designed to position the molecules at the receiver site with a high density for as long as possible to maximize the probability of reception/detection. In this way, we were able to increase the expected number of observed molecules and improve the BER performance. Our work is based on the assumption that molecule velocity is proportional to the electric field strength.  We verified this assumption with an analysis of the fluid dynamics of the system.  Analytical and numerical results demonstrated the feasibility and efficacy of the proposed electrophoretic approach to achieving molecular communication.  Given the compelling results from this investigation, future efforts will be focused on refining the model to confined propagation media with a view to exploring the suitability of these techniques in specific lab-on-a-chip applications.

\section*{Funding acknowledgement}

This work was made possible by the OUP's John Fell Fund (Grant No. 0006235). AACP and TCS were also funded by a Royal Society University Research Fellowship (Grant No. URF\textbackslash R\textbackslash 180016) and Enhancement Award (Grant No. RGF\textbackslash EA\textbackslash 181002).

\bibliographystyle{IEEEtran}
\bibliography{reference.bib}

\end{document}